\documentstyle[11pt,paspconf,epsfig]{article}
\begin{document}

\title{THE NGST  AND THE ZODIACAL LIGHT\\ IN THE SOLAR SYSTEM}

\author{Nick Gorkavyi\altaffilmark{1}, Leonid Ozernoy\altaffilmark{2},
John Mather}
\affil{NASA Goddard Space Flight Center}

\author{and Tanya Taidakova}
\affil{Computational Consulting Service}

\altaffiltext{1}{NRC/NAS Research Associate}
\altaffiltext{2}{Physics \& Astronomy Dept. and Inst. for Computational
 Sciences, George Mason University }

\begin{abstract}
We develop a physical model of the zodiacal cloud incorporating the 
real dust sources  of asteroidal, cometary, and kuiperoidal origin.
Using the inferred distribution of the
zodiacal dust, we compute its thermal emission and scattering at several 
wavelengths (1.25, 5, and 20 $\mu$m) as a function of NGST location assumed 
to be at 1 AU or 3 AU. Areas on the sky with a minimum of zodiacal light are 
determined.
\end{abstract}

\section{Physical model for zodiacal thermal radiation and scattering} 

Improvements in the zodiacal light emission and scattering to be given
by space observations at 3 AU, compared
with observations near the Earth, were discussed by {\it Mather and Beichman
 (1996)}. Unfortunately, a rather accurate multi-parametric model of the
zodiacal brightness derived by {\it Kelsall et al. (1998)} from the COBE data
cannot be reliably extrapolated to heliocentric distances as large as 
3 AU. We have  developed a physical model of the zodiacal cloud incorporating 
the real dust sources  of asteroidal and cometary origin, which 
makes it possible to evaluate quantitatively the zodiacal light emission 
and scattering throughout the Solar system ({\it Gorkavyi et al. 1997a}).
This model  considerably improves our previous `reference model' based
on the use of the continuity equation for distribution function of dust 
particles ({\it Gorkavyi et al. 1997b,c, 1998}) and enables us to
obtain more reliable results than a phenomenological modelling of 
the zodiacal light (e.g. {\it Ebbets 1998}). Below, we describe an improved 
model that represents a 3D-grid containing $45\times 180\times  244=2\cdot 
10^6$ cells with a  step in (heliocentric latitude $\varphi$, longitude 
$\lambda$, and radius $R$) to be $(2^\circ, 2^\circ, ~0.025~R$ [AU]).
Using a new numerical approach to the dynamics of minor bodies and
 dust particles, we increase the number of particle positions employed in 
each model to $10^{11}$ without using a supercomputer 
({\it Ozernoy et al. 2000, Gorkavyi et al. 2000}). We compute here 
the distribution of the zodiacal dust emission 
and scattering in the Solar system. The processes 
influencing the dust particle dynamics include gravitational scattering on
all planets, except Pluto (inclinations and precession of the planets
are neglected), mean motion resonances, and the
Poynting-Robertson/solar wind drags.

We employ here 931 sources of dust particles (with eccentricities $e<1$), 
which include 284 asteroids, 451 short-period comets, and 196 Kuiper belt 
objects.
We adopt a two-component approximation for   dust particle size distribution, 
which is characterized by the parameter $\beta=0.285$ and 0.057 for
(1-2)$\mu$m and (5-10)$\mu$m particles, respectively.

 The inferred dust density distribution, $n(R,\varphi,\lambda)$,
of the IPD at heliocentric distances $0.5<R<100$ AU
enables us to compute
the scattered light and the thermal emission of
the zodiacal cloud as a function of the observer's latitude and
longitude.

The thermal emission and scattering of the zodiacal cloud are given by:
$$I_\nu \propto \int n(R, \varphi,\lambda) \left[(1-A)B_\nu(T) + 
A F_\nu^\odot \Phi(\Theta)\right]ds$$
where $B_\nu$ is the Planck function at frequency $\nu$, $A$ is albedo,
$T$ is the dust temperature, $T(R)=T_0R^{-\delta}$ with $T_0=286^\circ$K
and $\delta=0.467$ ({\it Kelsall et al. 1998}),
$F_\nu^\odot \propto 1/R^2$ is the solar flux,
$\Phi(\Theta)$ is the phase function at scattering angle $\Theta$ 
 ({\it Kelsall et al. 1998}).

Note that although the phase function only weakly depends on $\nu$,
the value of $I_\nu$ strongly depends on $\nu$ through the solar flux. Besides,
the Planck function strongly depends on heliocentric distance, especially 
in the near IR (1-5 $\mu$m).

\section{The zodiacal light at R=1 and 3 AU}

Using the inferred distribution of the zodiacal dust,
we have computed a variety of zodiacal light maps, both for thermal    
emission and scattered components, at different locations 
$(R,Z)$ of the NGST assumed to be at 1 AU or 3 AU (see Fig.~1 for the
$5\mu$m-emission of
asteroidal dust and Fig.~2 for that of kuiperoidal dust). The brightness
as a function of  latitude $\varphi$ and longitude $\lambda$ (in the
telescope's frame) is given in logarithmic scale (the neighboring contour
intensities differ by ${\sqrt e}$). The Sun's position is
(0,0) for $Z=0$ and is shifted to a negative  $\varphi$  for $Z=0.25$ AU.

At each location, there is a minimum in the zodiacal light which can be seen
as a `dark spot' (or several spots) in the computed Figures. The positions of 
those minima are explained by an interplay between dependencies of emissivity
upon density and temperature.

Our numerical results concerning the computed zodiacal light at different 
locations of NGST are summarized in Tables~1 and 2.
The data presented in Table~2 for $R=1$ AU, 2nd column can be compared with 
the COBE data $I(0^\circ,90^\circ)/I(90^\circ)=3.5$ (1.25$\mu$m band), $=2.3$
(5$\mu$m band), and $=3.3$ (20$\mu$m band). Interestingly, {\it 
it is the kuiperoidal component (k1) whose shape resembles most closely 
the profile of the zodiacal cloud}, which might imply that the role of 
kuiperoidal particles has been highly underestimated so far. 
\begin{table}[!h]
\caption{Components of the model zodiacal cloud.}  \label{tbl-1}
\begin{center}\scriptsize
\begin{tabular}{crrrrr}
Origin/Size & \multicolumn{1}{c}{\# of sources\tablenotemark{a}}
 & \# of particles  &    \# of particles  & \# of positions \\
of dust particles   &     &  {with $e<1$ at $t=0$~\tablenotemark{a}}  &   
{near Earth\tablenotemark{a}}   &   in our 3D-model \\
\tableline
& & & &\\
asteroidal/1$\mu$m (a1)  &  110+110  &   110+110 &  57+85  &   $7\cdot 10^9$\\
cometary/1$\mu$m (c1)  &  2128+112 &  276+112  &     73+78 &  $7\cdot 10^9$\\
kuiperoidal/1$\mu$m (k1) &  100+100 &  96+100  &  11+12 &  $ 6\cdot 10^{10}$\\
asteroidal/5$\mu$m  (a5) &  32+32  &   32+32 &   32+32 &   $2\cdot 10^{10}$\\
cometary/5$\mu$m (c5) &  61+3 &   60+3  &  40+2  &         $1\cdot 10^{10}$\\
\end{tabular}
\end{center}
\tablenotetext{a}{Pericentral start $+$ apocentral start} 
\end{table}

\begin{table}[!h]
\caption{Brightness of different components of the zodiacal cloud  $~~~$
\scriptsize 
assuming the intensity of each component near the Earth be
$I(\varphi =90^\circ, \lambda )=1$  at the pole}  \label{tbl-2}
\begin{center}\scriptsize 
~~~~~~~~$R=1$ AU, $Z=0$ AU~~~~~     ~~~~~~~~~$R=1$ AU, $Z=0.25$ AU~~~~~~~ 
~~~~~~~~~~$R=3$ AU, $Z=0$ AU~~~~~~~\\
\begin{tabular}{crrrrrrrrrrrrrrr}
\multicolumn{1}{c}{~0\tablenotemark{*}~~} 
&\multicolumn{1}{c}{~~2\tablenotemark{b}~~~}
&\multicolumn{1}{c}{~3\tablenotemark{c}~~~}
&\multicolumn{1}{c}{4\tablenotemark{d}~~~}
&\multicolumn{1}{c}{5\tablenotemark{e}~~~~~~}
&{1\tablenotemark{a}~~}
&{2\tablenotemark{b}~~~~}
&{3\tablenotemark{c}~~}
&{4\tablenotemark{d}~~~}
&{5\tablenotemark{e}~~~~~}
&{1\tablenotemark{a}~~~~~}
&{2\tablenotemark{b}~~}
&{3\tablenotemark{c}~~~}
&{4\tablenotemark{d}~~~}
&{5\tablenotemark{e}~~}\\
\tableline
\end{tabular}
\begin{tabular}{cr}
\multicolumn{1}{c}
{}\\
{$1.25~\mu$m band}\\
\end{tabular}
\begin{tabular}{crrrrrrrrrrrrrrr} 
a1 & 4.8 & 3.2 & .92 &  74
 &   ~~.20 &  1.7 & 1.3  & .20 &  87 &
~~.068 & .29 &  .11 &  .063 & 74\\
c1  & 1.9 & 1.2 & .91 &  75
 &  ~~.42 &  1.9 & 1.4 &  .42 &  89  &
~~.058 & .13 &  .083 & .052 & 54\\
k1  & 3.9 & 2.7 & .95 &  73
 & ~~.17 &  2.0 & 1.8 &  .16 &  89 &
~~.050 & .55 &  .60 &  .046 & 87\\
a5  & 7.3 & 5.0 & .90 &  74
 & ~~.073 & 1.1 & .95 & .073 & 88 &
~~.018 & .059 & .018 & .012 & 45\\
c5  & 1.6 & 1.0 & .87 &  30
 & ~~.54 &  1.5 & 1.0 &  .55 &  89 &
~~.045 & .068 & .046 & .035 & 57\\
&
&
&
&
&
&
&
&
&
&
&
&
&
& \\
\end{tabular}
\begin{tabular}{cr}
\multicolumn{1}{c}{$5~\mu$m band\tablenotemark{f}}\\
\end{tabular}
\begin{tabular}{crrrrrrrrrrrrrrr} 
a1 &  2.9 & 1.0 &  .79 & 41
 & ~~.18 &  .79 & .31 & .16 &  82 &
~~~~.69 & ~~1.8 &  ~~.49 &   ~.46 &  38\\
c1 &  1.3 & .44 & .44 &  6
 &  ~~.38 &  1.2 &  .41 & .32 &  64 &
~~~~.53 &  ~~.67 &  ~~.20 &   ~.20 &   8\\
k1 &  2.3 & .76 & .74 & 39
 &  ~~.16 &  .80 & .29 & .15 &  89 &
~~~~.44 &  ~~.72 &  ~~.20 &   ~.20 &  10\\
a5 &  4.9 & 1.8 &  .87 & 67
 &  ~~.066 & .52 & .22 & .061 & 81 &
~~~~.19 &  ~~.54 &  ~~.11 &   ~.11 &  45\\
c5 &  1.2 & .41 & .40 & 14
 &  ~~.49 &  1.1 &  .36 & .30 &  14 &
~~~~.42 &  ~~.49 &  ~~.13 &   ~.13 &   8\\
&
&
&
&
&
&
&
&
&
&
&
&
&
& \\
\end{tabular}
\begin{tabular}{cr}
\multicolumn{1}{c}{$20\mu$m band}\\
\end{tabular}
\begin{tabular}{crrrrrrrrrrrrrrr} 
a1 &  5.2 & 2.7 &  .97 & 74 &  ~~.22 &  1.9 & 1.2 &  .21 &   
82 & ~~.10 &  .44 &  .18 &  .097 & 74\\
c1 &  2.0 & 1.0 &  .95 & 73 &  ~~.46 &  2.0 & 1.1 &  .45 &   
89 & ~~.083 & .18 &  .084 & .068 & 42\\
k1 &  3.8 & 1.9 &  .95 & 83 &  ~~.18 &  1.9 & 1.1 &  .17 &   
89 & ~~.069 & .35 &  .23 &  .064 & 79\\
a5 &  8.7 & 4.4 &  .99 & 83&  ~~.086 & 1.4 & .83 & .085 &  
88 & ~~.029 & .095 & .023 & .018 & 45\\
c5 &  1.7 & .81 & .77 & 16 & ~~.59 &  1.6 & .77 & .59 &   
89 & ~~.066 & .096 &  .039 & .039 &  7\\
\end{tabular}
\end{center}
\tablenotetext{*}{Type of dust listed in Table~1.}
\tablenotetext{a}{intensity at the north pole, $I_{pole}$}
\tablenotetext{b}{intensity at the ecliptic plane in the direction of 
$90^\circ$ from the Sun, $I(\varphi=0^\circ,~\lambda=90^\circ)$}
\tablenotetext{c}{intensity at the ecliptic plane in anti-sun direction, 
$I(\varphi=0^\circ,~\lambda=180^\circ)$}
\tablenotetext{d}{$I_{\rm min}(\varphi_{\rm min},~\lambda=180^\circ)$, where
$\varphi_{\rm min}$ is the latitude at which the zodiacal light brightness\\
is minimal in the anti-sun direction ($\lambda=180^\circ$)}
\tablenotetext{e}{$\varphi_{\rm min}$ [$^\circ$]}
\tablenotetext{f}{All $5\mu$m-intensities at $R=3$ AU, $Z=0$ need to be 
multiplied by $10^{-3}$}
\end{table}

\section{Conclusions}

1. The structure of the zodiacal dust cloud computed in the present work
is substantially non-uniform:
 
   (i) near the Earth,  the thickness of the dust layer is the largest for
cometary particles and the smallest for asteroidal particles, with
 kuiperoidal particles occupying an intermediate position;

    (ii) the larger the size of asteroidal dust particles, the thinner is the 
layer comprised of such particles; for cometary particles a reverse, although
a weaker, dependence takes place;

    (iii) the partial contribution 
of particles of  different origin and size changes with heliocentric
distance: as the distance increasess, the contribution of asteroidal particles 
(especially of large size) into the zodiacal light emission sharply 
decreases, whereas the contribution of the kuiperoidal component increases;

(iv) The latitudinal dependence of the zodiacal  emission is  different for 
different components of the dust cloud. In the anti-sun direction ($\lambda=
180^\circ$), the maximum of the asteroidal dust emission at 
$5~\mu$m is reached at the ecliptic plane, whereas the 
brightness of the cometary component has its minimum here. 

2. Observations made from $R=1$ AU, $Z=0.25$ AU
 at all wavelengths  would detect  the brightness
 of the zodiacal light in the direction of the pole
 at level of 20-30\% of that seen near the Earth. 

3. At $R=3$ AU, the zodiacal light brightness 
in the direction of the pole decreases  to
$5-6$\%  at $1.25~\mu$m;
 to $0.05-0.07$\% at $5~\mu$m, and to  $8-10$\% 
at $20~\mu$m compared to that seen near the Earth.

4. At each wavelength, there are certain regions on the sky having  a minimal
zodiacal light brightness. At short wavelengths ($\sim 1.25~\mu$m),
these regions are around the poles, and at long wavelengths (such as
5 to 20 $\mu$m) these regions are shifted toward the ecliptic. This result 
needs to be taken into consideration while planning observations with NGST. 

5. Further improvements in our physical modelling of the zodiacal light
are feasible. Using COBE and IRAS data, the contribution of dust components
of different origin into the zodiacal cloud could be determined. Those 
improvements would make far going implications possible for 
extragalactic astronomy and cosmology.

\acknowledgements This work has been supported by NASA grant NAG5-7065
to George Mason University. N.G. acknowledges the NRC-NAS associateship.

\vspace{0.1truein}

\noindent{\bf References}
\def\ref#1  {\noindent \hangindent=24.0pt \hangafter=1 {#1} \par}
\def\v#1  {{\bf {#1}{\rm,}\ }}
\def\ul#1{$\underline{\smash{\vphantom{y}\hbox{#1}}}$}
\vspace{0.08truein}

\ref{Ebbets, D. 1998, ``Zodical Light Model for  NGST Studies". 
(unpublished)}
\ref{Gorkavyi, N.N., Ozernoy, L.M., \& Mather, J.C. 
1997a, ApJ 474, 496}
\ref{Gorkavyi, N.N., Ozernoy, L.M., Mather, J.C. \& Taidakova, T.A.
1997b, BAAS 29, 1310}
\ref{--------
1997c, ApJ 488, 268}
\ref{--------
1998, Earth, Planets and Space, 50, 539}
\ref{--------
2000, Planetary and Space Science (submitted)}
\ref{Kelsall, T. et al. 1998, ApJ 508, 44}
\ref{Mather, J.C. \& Beichman, C.A. 1996, ``Unveiling the Cosmic Infrared 
Background", Ed. E.~Dwek, p.~271}
\ref{Ozernoy, L.M., Gorkavyi, N.N. \& Taidakova, T.A. 2000, Icarus (submitted)}

\begin{figure}[t] 
\centerline{\epsfig{file=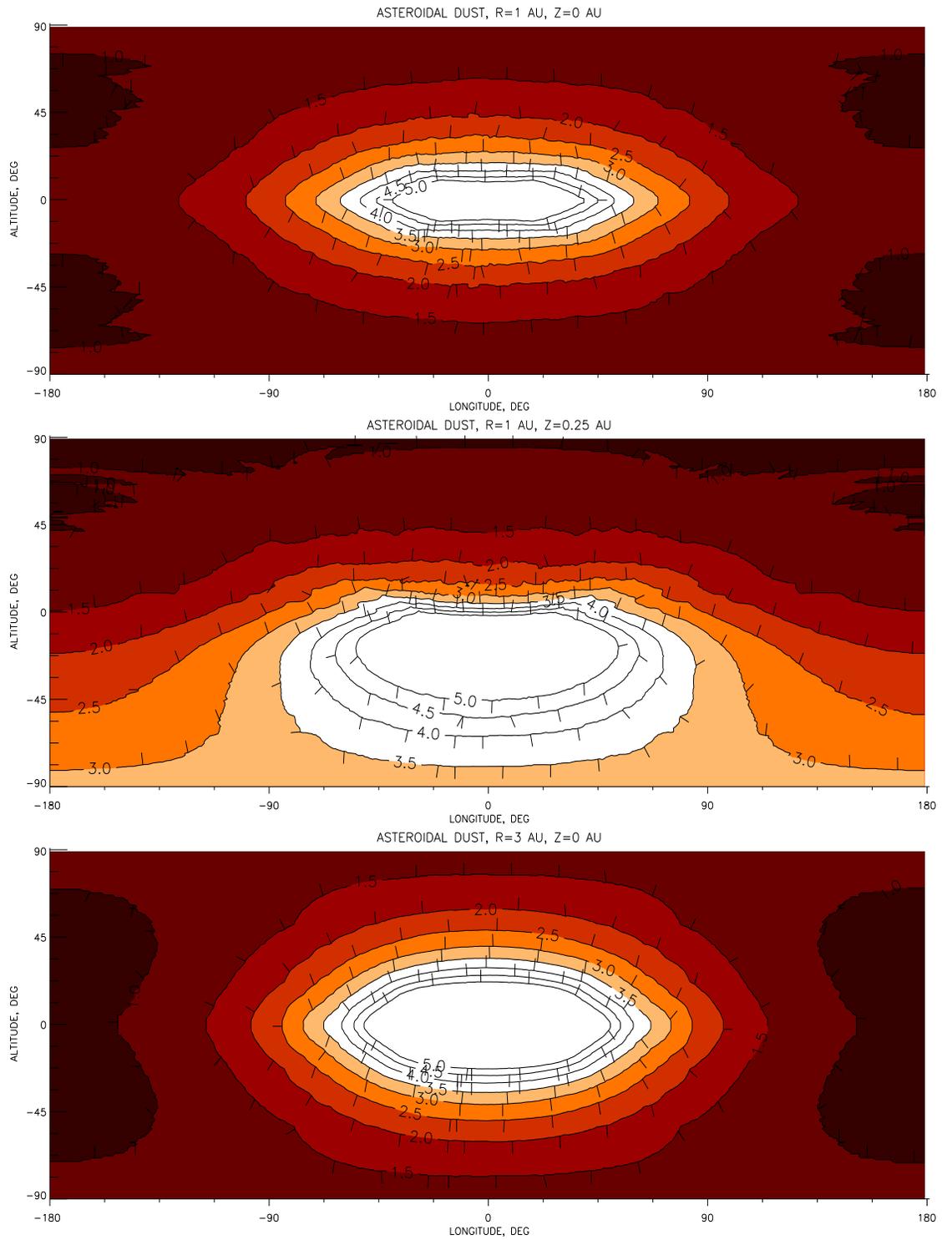,width=6.0in,height=7.75in}}
\caption{Emission of a small ($r=1-2\mu$m) asteroidal dust in the 5$\mu$m band
}
\label{fig1}
\end{figure}
\begin{figure}[t] 
\centerline{\epsfig{file=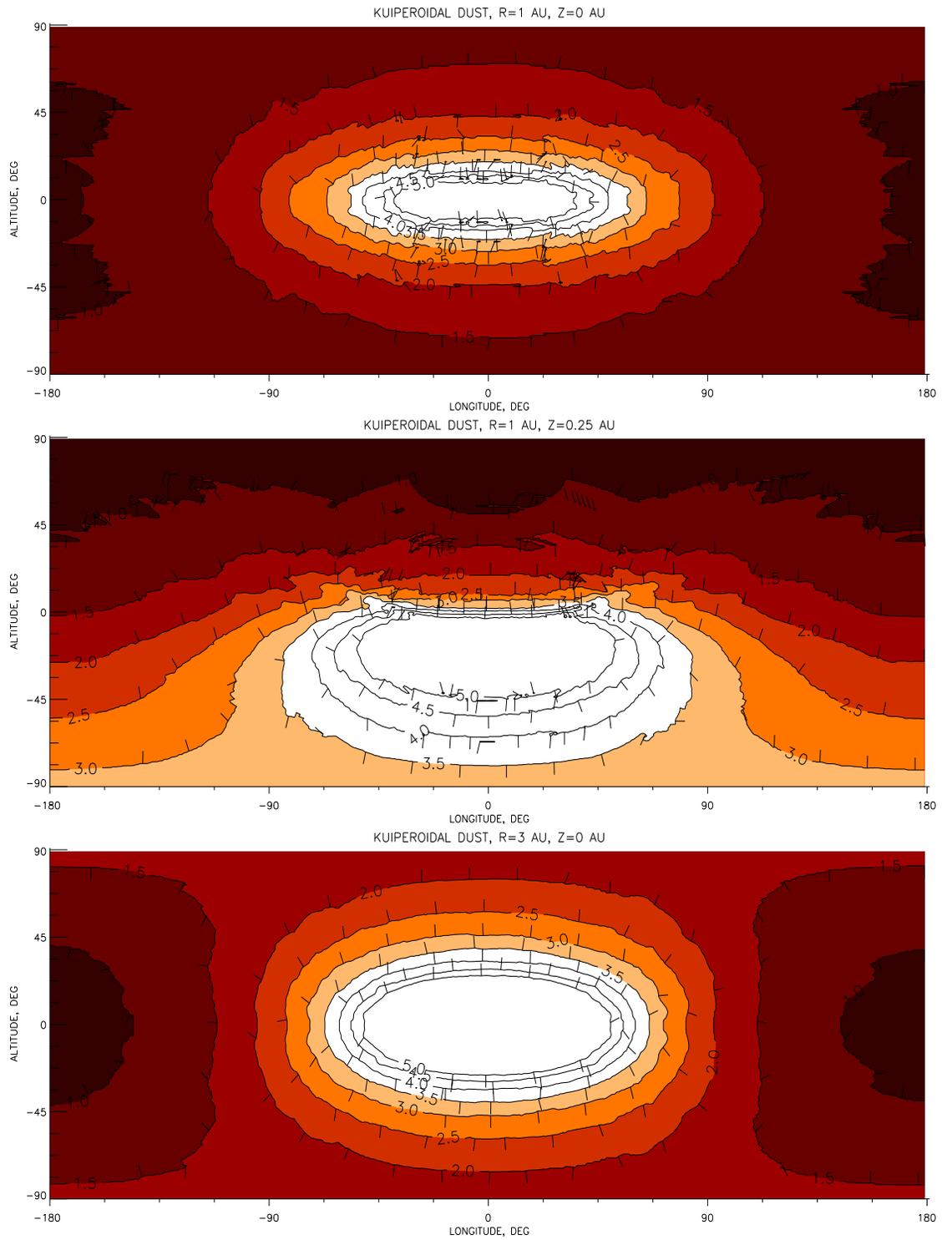,width=6.0in,height=7.75in}}
\caption{Emission of a small ($r=1-2\mu$m) kuiperoidal dust in the 5$\mu$m band
}
\label{fig2}
\end{figure}
\end{document}